# Study of detailed balance between excitons and free carriers in pristine diamond using terahertz spectroscopy

T. Ichii,[1, a)] Y. Hazama,[1] N. Naka,[1] and K. Tanaka[1, 2, a)]

[1]*Department of Physics, Department of Physics, Kyoto University, Kyoto 606-8502, Japan*

[2]*Institute for Integrated Cell-Material Sciences (WPI-iCeMS), Kyoto University, Kyoto 606-8501, Japan*

[a)] *Corresponding author: Electronic mail: kochan@scphys.kyoto-u.ac.jp and ichii.tomoaki.46v@st.kyoto-u.ac.jp*

A fundamental understanding of the photoexcited carrier system in diamond is crucial to facilitate its application in photonic and electronic devices. Here, we studied the detailed balance between free carriers and excitons in pristine diamond by using a deep-ultraviolet (DUV) pump in combination with terahertz (THz) probe spectroscopy. We investigated the transformation of photoexcited carriers to excitons via an internal transition of excitons, which was newly found to occur at a frequency of approximately 15 THz. We determined the equilibrium constant in the Saha equation from the temperature dependence of the free-carrier density measured at chemical equilibrium. The derived exciton binding energy was larger than the conventional value, which indicated an energy shift due to the fine-structure splitting of the exciton states.

Compared to conventional semiconductors such as Si, Ge, and GaAs, diamond has exceptional electronic properties such as a large band gap, thermal conductivity, and breakdown voltage [1,2]. Because of these beneficial properties, diamond has applications in power devices and high-energy particle detectors [3–7]. Diamond also has a large exciton binding energy of ~80 meV (930 K) [8,9], which leads to significant formation of excitons even at room temperature. As a result, it has potential applications in devices such as deep-ultraviolet (DUV) light-emitting diodes, which make use of exciton luminescence [10]. Despite the necessity for more practical and efficient use of such diamond-based devices, the detailed balance between excitons and free carriers remains a subject for debate.

The chemical equilibrium between excitons and free carriers in semiconductors can be described using the Saha equation, $n_{eh}^2/n_{ex} = A \cdot T^{3/2}\exp(-E_{ex}/k_B T)$ [11,12], where the exciton binding energy $E_{ex}$ enters through the Boltzmann factor. For a given coefficient $A$ and temperature $T$, the exciton density $n_{ex}$ and free-carrier density $n_{eh}$ can be predicted at various total carrier densities. Theoretically, the value of $A$ for diamond [13] is estimated from the effective masses of free carriers and excitons to be on the order of $10^{14}$ cm$^{-3}$ K$^{-3/2}$. However, the experimental values of $A$ lie in a wider range of $10^{12}$–$10^{14}$ cm$^{-3}$ K$^{-3/2}$ [13–16]. This large difference between experimentally determined and theoretically calculated $A$ values is attributable to several complicated phenomena occurring in the photoexcited carrier system, such as diffusion, the Auger process, and surface recombination [14–16]. Therefore, there is a need for a new method for an accurate determination of the coefficient $A$.

Here, we propose the use of terahertz time-domain spectroscopy (THz-TDS) [17–19] for the estimation of $A$. The THz-TDS technique facilitates determination of the absolute densities of photoexcited carriers and excitons from the complex dielectric function [20] and therefore provides the exciton ionization ratios in Si, Ge, and GaAs quantum wells [20–22]. However, we had to overcome two technical challenges before we could apply the THz-TDS technique to diamond. The first challenge was the generation and detection of a THz pulse that covers the excitonic internal transition region, which is expected to be at approximately 15 THz. For example, the accessible spectral region of conventional THz-TDS systems based on nonlinear crystals is typically limited (2–5 THz) owing to phonon absorption and dispersion [18-23]. To overcome this challenge, we used air as the medium for the THz emitter [24] and detector [25] owing to its low dispersion. The second challenge was the lack of transparent windows in both the DUV region of the optical pump and the THz frequency region of the probe. To overcome this challenge, we custom-designed a cryostat with three exchangeable windows made of different materials (Ge, Si, and quartz). Following these technical improvements, we estimated the coefficient $A$ of diamond to be $(4.4 \pm 2.7) \times 10^{14}$ cm$^{-3}$ K$^{-3/2}$ from the temperature dependence of the free-carrier density by a DUV pump–THz probe spectroscopy system. Our results provided an accurate detailed-balance equation, which is expected to be useful in the design of novel applications based on diamond.

We used chemical-vapor-deposition-grown diamond (Element Six) with nitrogen (<5 ppb) and boron (1 ppb) as unintentionally incorporated dopants. The thickness of the sample was 500 μm. Figure 1 shows the experimental setup of the DUV pump–THz probe spectroscopy system. An 800 nm laser pulse generated from a Ti:sapphire amplifier was split into three beams: one for generating the DUV pulse, one for generating the THz pulse, and one for detecting the THz field. The DUV pulse (267 nm) was produced via third harmonic generation (THG) [26] and was incident on the sample at an angle of 45°. The spot diameter and excitation density were 0.7 mm and 2.5 mJ/cm$^2$, respectively. According to Snell's law [27], light propagated with a spot size of 1.0 mm at an inside angle of 19°. The DUV pulse homogeneously generated transient free carriers throughout the sample with $n_{eh}$ on the order of $10^{15}$ cm$^{-3}$ via a two-photon absorption process [26]; this value is lower than the critical density of the exciton Mott transition ($4 \times 10^{18}$ cm$^{-3}$ as determined using the Thomas–Fermi screening approximation [28]). The broadband THz pulse (2–25 THz) was generated by a collinear air plasma method [24] and detected by the air-biased coherent detection (ABCD) method [25]. The spot diameter of the THz pulse as measured using a THz camera (NEC IRV-T0831) was 140 μm. By transmitting the THz pulse through the sample with and without the DUV pulse, we obtained the photoinduced change in the complex dielectric function [18].

We studied the dynamics of exciton formation by observing their internal transition (1s-2p transition), which is expected to occur at $\hbar\omega_{ex} \sim 15$ THz when excitons with $E_{ex} = 80$ meV [8,9] have Rydberg energy levels. The cryostat windows for the THz pulse were made of Ge. Figure 2(a) shows the changes in the complex dielectric function $\Delta\varepsilon$ (=$\Delta\varepsilon_1 + i\,\Delta\varepsilon_2$) as measured under the application of the DUV pulse at three time delays ($\Delta t$ = -10, 10, and 550 ps) at a lattice temperature ($T_L$) of 100 K. We observed the Drude response at $\Delta t = 10$ ps, which indicated the generation of free carriers by the DUV pulse. At $\Delta t = 550$ ps, the Drude component decreased with an increase in the peak structure at approximately 15 THz, which was attributed to the internal transition of excitons. We analyzed the measured data with two spectral components by employing Drude and Lorentz oscillator models:

$$\Delta\varepsilon(\omega) = -\frac{n_{eh}e^2}{\varepsilon_0 m^*}\frac{1}{\omega(\omega+i\gamma_{eh})} - \frac{n_{ex}e^2}{\varepsilon_0 m^*}\frac{1}{\omega^2-\omega_{ex}^2+i\omega\gamma_{ex}} \quad (1)$$

where $\varepsilon_0$ is the vacuum permittivity and $m^* = (1/m_e + 1/m_h)^{-1} = 0.19m_0$ [29] is the reduced mass of the excitons in diamond. The spectra were fitted by varying the following parameters: $n_{eh}$, $n_{ex}$, the damping constants ($\gamma_{eh}$ and $\gamma_{ex}$), and the resonance frequency ($\omega_{ex}$). $\Delta\varepsilon$ at 10 ps was well reproduced by the Drude component alone, i.e., $n_{ex} = 0$ (blue solid line), and $n_{eh}$ was estimated to be $4.9 \times 10^{15}$ cm$^3$. This value is consistent with the value estimated from the DUV pulse energy. In addition, $\Delta\varepsilon$ at 550 ps was well reproduced by the Drude and Lorentz components (black solid line). $n_{ex}$ and $n_{eh}$ were estimated to be $4.1 \times 10^{15}$ cm$^{-3}$ and $0.49 \times 10^{15}$ cm$^{-3}$, respectively. The total density of $e$–$h$ pairs, $n_{total} = n_{ex}(t) + n_{eh}(t)$, at 550 ps was almost identical to the initial free-carrier density at 10 ps. This was confirmed from the time evolutions of $n_{eh}$, $n_{ex}$, and $n_{total}$, as shown in Figure 2(b). Therefore, the spectral change represented the transformation of free carriers to excitons with conservation of $n_{total}$. The conservation of $n_{total}$ indicated that this transformation occurred before the annihilation of the excitons and free carriers. The sub-nanosecond decay of $n_{eh}$ was considerably shorter than the lifetime of carriers in intrinsic diamond [30] owing to their Auger recombination (0.1–1.0 μs with $n_{eh}$ of approximately $10^{15}$ cm$^{-3}$) and trapping by impurities (0.1–1 μs). Under our experimental conditions, $n_{ex}(t)$ was estimated as $n_{total} - n_{eh}(t)$, where $n_{total}$ was defined as the initial value of $n_{eh}$. Therefore, in the next part of the study, we focused on $n_{eh}(t)$ extracted via spectral shape analysis of the Drude component at various delay times and lattice temperatures.

Figure 3(a) shows the time evolutions of the changes in the complex dielectric function at $T_L$ of 160 K and 100 K. The Ge windows of the cryostat were replaced with Si windows, which were transparent in the THz region (2–10 THz) of the Drude response curve. Both $\Delta\varepsilon_1$ and $\Delta\varepsilon_2$ were well reproduced by the Drude model (solid lines), and they yielded the time evolution of $n_{eh}(t)$ at these lattice temperatures (Figure 3(b)). At $T_L = 160$ K, $n_{eh}$ decreased rapidly with increasing time delay and attained a constant value after 100 ps. The carrier dynamics was found to be significantly dependent on the lattice temperature; a longer decay of approximately 200 ps and smaller constant value of $n_{eh}$ were observed at $T_L = 100$ K.

The observed sub-nanosecond dynamics and temperature dependence indicated that the decay of $n_{eh}$ was presumably caused by the thermal relaxation of carriers in the lattice system, as previously reported for Si and GaAs quantum wells [20,22]. To confirm the thermalization of the photoexcited carriers in the lattice system within 200 ps, the cooling dynamics of the carrier temperature, $T_C(t)$, in diamond was analyzed numerically via calculation of all the relevant phonon relaxation rates. The average rate of energy loss per electron or hole due to acoustic and optical phonons has been described previously [31]. In the present study, the values of effective masses as parameters used in this calculation were determined via cyclotron resonance measurements [29]. The values of all other parameters, i.e., deformation potentials and phonon frequencies, are listed in the Supplementary Material. The resulting cooling curves at 160 K and 100 K are shown in Figure 3(c). The decay of time-dependent free-carrier densities obtained by the Drude fitting was well reproduced by the calculated curve of the carrier temperature; this result supports the presumption that the decay of $n_{eh}$ was due to exciton formation accompanied by the thermalization of free carriers within the lattice system.

We further measured the changes in the complex dielectric function with the lattice temperature at $\Delta t = 550$ ps where the cooling dynamics of photoexcited carriers are complete. Fitting analysis using the Drude model enabled determination of the free-carrier density at each lattice temperature $T_L$. The obtained $n_{eh}$ is represented by the dots in Figure 4 as a function of $T_L$. We analyzed the temperature dependence of the free-carrier density by employing the Saha equation [20–22], which describes the densities of excitons and free carriers after statistical equilibration of their chemical potentials in the classical limit. The Saha equation for a classical three-dimensional gas of free $e$-$h$ pairs and excitons is given as

$$\frac{n_{eh}^2}{n_{ex}} = A\, T_L^{\frac{3}{2}} \exp\left(-\frac{E_{ex}}{k_B T_L}\right), \quad A = \left(\frac{\mu k_B}{2\pi\hbar^2}\right)^{\frac{3}{2}} \qquad (2)$$

where $E_{ex}$ is the exciton binding energy and $\mu = m_{de}m_{dh}/(m_{de} + m_{dh})$. Here, $m_{de}$ and $m_{dh}$ are the density-of-state (effective) masses of electrons and holes [21], respectively. By taking $n_{total}$ as $1.06 \times 10^{15}$ cm$^{-3}$, which is the initial free-carrier density at $\Delta t = 10$ ps and 160 K, we obtained the best-fit values of $A$ and $E_{ex}$. From the black solid line in Figure 4, we obtained $A = (4.4 \pm 2.7) \times 10^{14}$ cm$^{-3}$ K$^{-3/2}$ and $E_{ex} = 93.8 \pm 8.2$ meV.

The value of $A$ could be estimated from the effective masses of electrons and holes. The theoretical value of $A$, i.e., $A_{th} = 4.5 \times 10^{14}$ cm$^{-3}$ K$^{-3/2}$, was obtained using $m_{de} = 0.39m_0$ and $m_{dh} = 0.94m_0$, which were directly determined via cyclotron resonance measurements [29]. Our experimental value of $A_{exp} = (4.4 \pm 2.7) \times 10^{14}$ cm$^{-3}$ K$^{-3/2}$ is consistent with this value. It should also be noted that the estimated $E_{ex} = 93.8 \pm 8.2$ meV is larger than the conventional value of 80 meV [8,9]. Because the 1s states of excitons in diamond undergo fine-structure splitting with a magnitude on the order of 10 meV [32], the obtained value of $E_{ex}$ is expected to be related to the binding energy of the lowest-energy excitons at the 1s level. Further details will be reported in our forthcoming paper on analysis of internal transition spectra of excitons by THz-TDS. Finally, we compared our results with those obtained using Saha equations adopted in previous studies [13–16]. As is clear from the dotted lines in Figure 4, the previous results deviated significantly from our results. This difference was due to the difference in not only $A$ (or effective masses) but also $E_{ex}$, which relied on the value that was determined at a higher temperature without considering the fine structure of exciton.

In summary, we developed an experimental DUV pump–THz probe spectroscopy system for pristine diamond and evaluated the temperature dependence of the free-carrier density under chemical equilibrium. We successfully determined the equilibrium constant in the Saha equation experimentally and confirmed its agreement with the theoretically estimated value. We also determined an additional value of the exciton binding energy ($E_{ex} = 93.8 \pm 8.2$ meV) as an alternative to the conventional value of 80 meV. These fundamental findings are expected to contribute significantly to the design of diamond-based photonic and electronic devices.

**Acknowledgement**
This work was supported by a Grant-in-Aid for Scientific Research (S) (Grant No. 17H06124), JST ACCEL (Grant No. JPMJMI17F2), a Grant-in-Aid for Scientific Research (B) (Grant No. 17H02910), and a Grant-in-Aid for Challenging Research (Exploratory) (Grant No. 19K21849) from JSPS, Japan.

**Data availability**
The data that support the findings of this study are available from the corresponding author upon reasonable request.

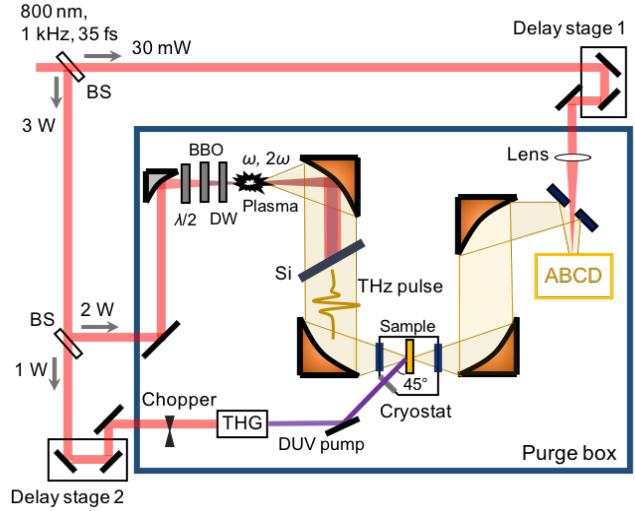

Figure 1. Schematic of experimental setup of DUV pump–THz probe spectroscopy system. BS: beam splitter; DW: dual-wavelength wave plate. The THz wave was generated by mixing a polarization-controlled fundamental pulse ($\omega$) through the DW and $\lambda/2$ plates and its second harmonic generation ($2\omega$) from a $\beta$-barium borate (BBO) crystal near the air plasma point. The THz wave transmitted through the sample was detected using a commercially available ABCD system (Zomega, ZAP-APD). The cryostat windows for the DUV and THz pulses were made of quartz and Si/Ge, respectively.

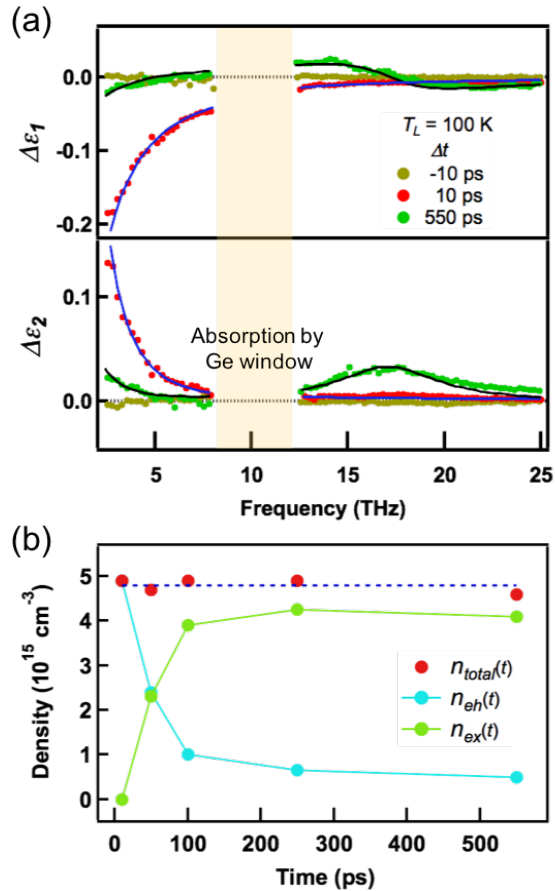

Figure 2. (a) Transient change in complex dielectric function at lattice temperature $T_L$ of 100 K after DUV-pump excitation. The spectral region of 7–13 THz is omitted because of the strong absorption by the Ge window of the cryostat. (b) Time evolutions of $n_{ex}(t)$ and $n_{eh}(t)$ estimated by Drude response and Lorentz model, respectively. $n_{total}(t)$ is the sum of $n_{ex}(t)$ and $n_{eh}(t)$. The blue dashed line is a guide to the eye.

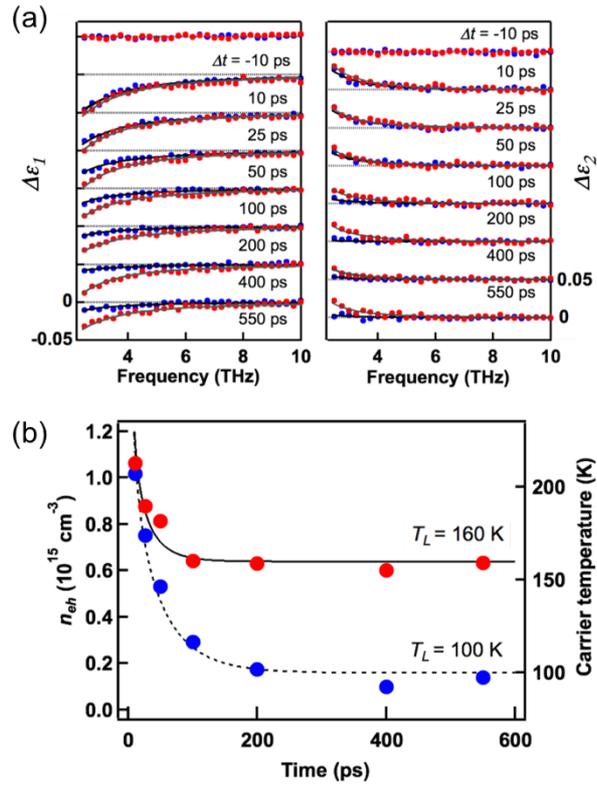

Figure 3. (a) Transient changes in complex dielectric function at lattice temperatures ($T_L$) of 100 K (blue dots) and 160 K (red dots) after DUV-pump excitation with Si window of cryostat. The solid lines represent the results of fitting using the Drude model. (b) Temporal free-carrier density determined by Drude fitting (represented by dots). The solid and dashed lines represent the numerically calculated results of $T_C(t)$ at $T_L$ = 160 K and 100 K, respectively.

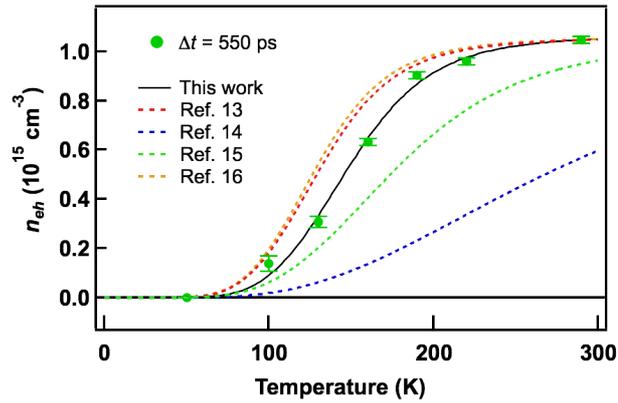

Figure 4. Temperature dependence of free-carrier density $n_{eh}$ ($\Delta t$ = 550 ps). The black solid line represents the fitting result of $n_{total}$ as obtained using the Saha equation. The dashed lines represent results of calculations using parameters adopted in previous studies. The values of coefficient $A$ in Refs. 13, 14, 15, and 16 were $4.5 \times 10^{14}$, $5.1 \times 10^{13}$, $3.6 \times 10^{12}$, and $5.3 \times 10^{14}$ cm$^{-3}$ K$^{-3/2}$, respectively. The value of $E_{ex}$ was 80 meV in all studies except for the full line.